# Adaptive Sampling of Dynamic Systems for Generation of Fast and Accurate Surrogate Models

Torben Talis[a], Joris Weigert[a,*], Dr. Erik Esche[a], Prof. Dr. Jens-Uwe Repke[a]

[a]Technische Universität Berlin, Process Dynamics and Operations Group, Sekr. KWT 9, Straße des 17. Juni 135, Berlin 10623, Germany

T.T. and J.W. contributed equally to this contribution

**Abstract:**

For economic nonlinear model predictive control and dynamic real-time optimization fast and accurate models are necessary. Consequently, the use of dynamic surrogate models to mimic complex rigorous models is increasingly coming into focus. For dynamic systems, the focus so far had been on identifying a system's behavior surrounding a steady-state operation point. In this contribution, we propose a novel methodology to adaptively sample rigorous dynamic process models to generate a dataset for building dynamic surrogate models. The goal of the developed algorithm is to cover an as large as possible area of the feasible region of the original model. To demonstrate the performance of the presented framework it is applied on a dynamic model of a chlor-alkali electrolysis.

## 1. Motivation & Introduction

The need for online reoptimization of continuously operated chemical plants becomes ever more important given the increase in demand response activity of industry, increases in feed fluctuations, or changes in demand, etc. (1). For processes with complex dynamics and slow return to steady-state, economic nonlinear model predictive control or dynamic real-time optimization has long been investigated (2, 3). Apart from the necessity to have highly accurate process models and reliable state estimators, fast and robust solution of the associated optimization problems is of the essence.

Hence, many research groups have started working on dynamic surrogate models, which accurately mimic the behavior of complex rigorous models of chemical processes and allow for fast computation of both state estimation and real-time optimization problems (4).

In these schemes, simulation problems using rigorous models are carried out offline and their results are then employed to train, e.g., recurrent neural networks, for online application (5). In these settings, the amount of simulations performed offline does not need to be limited. Rather, it is important that the simulations cover a large swath of the original model's feasible region in terms of both inputs (controls and initial conditions) and outputs (state variables) as most surrogate models have no guarantees regarding extrapolation.

Sampling and surrogate modeling for steady-state systems is well established (6, 7). For dynamic systems, the focus so far had been on "system identification", i.e., identifying a system's behavior surrounding a steady-state operation point (8). These methods are in general not capable to generate surrogate models capable of mimicking the behavior of a chemical plant from start-up to shutdown and have only a small range of validity.

### 1.1. Objective

To fill this gap, the present contribution proposes a novel methodology to adaptively sample rigorous dynamic process models, with the goal of covering an as large as possible area of the feasible region of the original model.

$$f(\dot{x}, x, u, d, p, t) = 0 \qquad (1)$$

The systems of interest are defined by Eq. (1), wherein $f$ is a set of differential-algebraic equations (DAE), $x$ are state variables, $u$ control variables, $d$ disturbances, $p$ model parameters, and $t$ time. The goal is to describe $f$ by a surrogate model $g$, which predicts $x$ of the next time point ($t_{k+1}$) based on the values of $x$ and $u$ of the current time point ($t_k$).

As a starting point for the sampling of $f$, we shall limit ourselves to the realistic assumptions of only one known set of initial values ($x_0 = x(t_0), u_0$) and that upper and lower bound for all controls are known ($u^L \leq u \leq u^U$).

Based on this initial knowledge, we here aim to create a dataset for building $g$ from scratch. Given that this initial information does not contain any information on the extent of the feasible region of $f$, nor does it hold information of the systems time constants beyond the initial point ($x_0, u_0$). By consequence, the proposed method will have to both explore the space of state variables $x$ as well as investigate frequencies at which the system $f$ shows excitations, which is subsequently relevant to determine the minimum time step for $g$.

## 2. State of the Art

For system identification, step experiments and oscillating input signals can be used for simple systems. These perturb a process at steady-state and generate data, which can be used to approximate the process by surrogate models valid in a limited area surrounding the steady-state operation point (9). In case of more complex systems, the choice of excitation signal is paramount. Multisine (10) as well as chirp (11) and amplitude modulated pseudo random binary signals (APRBS), which "can be understood as a sequence of step functions" (12), need to be tailored depending on the system's characteristics, i.e., delays, nonlinearity, time constants, etc. APRBS combines highly dynamic steps and low dynamic constant parts and covers the whole input space (11). Design of experiments may be used to maximize the information that can be achieved with every (simulation) experiment of the process (12). These methods typically focus on excitation of the system by manipulating $u$ and to hence generate data for $x$, while always starting from the same initial point $x_0$.

Naturally, this does not necessarily induce a large coverage of the feasible area in $x$. Many different methods are available to sample in hypercubes. Distributing points evenly in a $k$-dimensional hypercube can be achieved by a uniform grid. However, it requires an exponentially growing number of sample points with an increase in $k$. Non-uniform sampling techniques, such as Latin Hypercube (13), Hammersley Sequence (14), and Sobol (15), are more efficient, but cannot avoid the exponential growth in terms of required number of points. Halton and Hammersley sequences are used to generate well distributed, space-filling samples even in higher dimensions. Both are deterministic and every subsequence has the same space-filling properties (16). "Hammersley points are an optimal design for placing $n$ points on a $k$-dimensional hypercube" (17).

Applying these to generate different initial points for $x_0$, however, is ill-advised as these will almost certainly lead to infeasibilities. Given the complexities of sampling both in steady-state and dynamic systems, many different sampling methods have been developed for surrogate model creation. "One shot approaches" generate all samples at once, without incorporating any prior knowledge of the system. They provide a good coverage of the input space (18).

Adaptive sampling methods for static systems have recently become popular. They can be divided into exploration- and exploitation-based methods. The former try to obtain a wide coverage of the input space, while exploitation-based methods are driven by the training progress of the model. The latter require multiple iterations of model training.

In (7) an exploration-based method is proposed that estimates the feasible region in parameter space by using a predetermined number of samples. "Automated learning of algebraic models for optimization" (ALAMO) can be used to sequentially sample data and structurally improve the surrogate model of algebraic systems.

An exploitation-based method is presented in (19): The input space is divided into regions, which are sampled independently. The model is trained and evaluated on those regions. New samples are added to the region with the highest model error, improving the prediction.

A different method is proposed in (20). It combines exploration and exploitation and reduces the number of function evaluations. However, multiple surrogate models on different subsets of data are trained. Another hybrid method is described in (18). The exploration criterion is based on a Voronoi tessellation in the input space, and the exploitation part uses local linear approximations of the objective function.

All of these methods are used for steady state models. Adaption to dynamic models and time series forecast is not easily possible. Olofsson et al. (2021) use design of dynamic experiments for model discrimination (21). The exploration-based methods focus on coverage of the input space, while the exploitation-based methods focus on minimizing the number of samples and function evaluations. Contrarily to that, our proposed method is based on coverage of the output space and minimizes training time.

3. **Proposed algorithm for adaptive sampling**

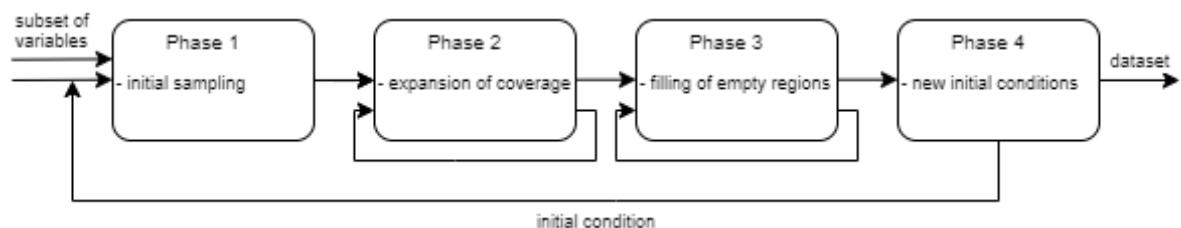

Figure 1. Proposed adaptive sampling method.

The proposed algorithm aims to generate a dataset for building a surrogate model. An overview is given in Figure 1. Multiple simulations with a short time horizon, a fixed timestep, and different inputs $\bar{u}$ are used to obtain a good coverage of the input space. (Bio-)chemical systems can have time constants differing by orders of magnitude. To identify these, a frequency modulated APRBS (FAPRBS) is proposed here and added on the inputs. It can be understood as a sequence of multiple

APRBS with different frequencies and is depicted in Figure 2. The maximum amplitude of the FAPRBS is small compared to the valid range of u.

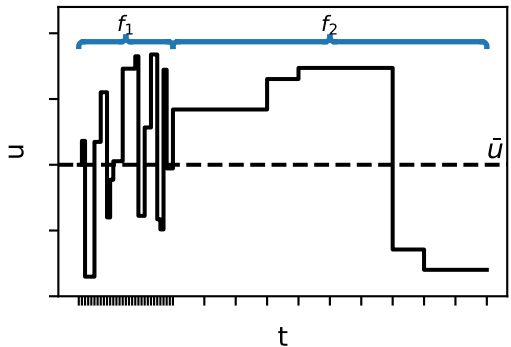

Figure 2. Frequency and Amplitude Modulated Pseudo-Random Binary Signal with 30 samples at $f_1$ and 10 samples at $f_2$. $f_1 \gg f_2$. The mean of the input signal $\bar{u}$ is given.

The overall algorithm is based on geometric quantities, especially the Euclidean distance of samples. The curse of dimensionality restricts the number of output variables which can be considered. A subset of all variables, that can contain state and non-state variables, must be selected. These variables form the output space $Y$. The dimensionality of $Y$ is currently limited to 7 by the applied implementation of the Quickhull algorithm (22, 23).

The trajectory of each simulation run will oscillate around a single point, which is called *seed* from here on. Based on the seeds, poorly covered areas in output space are identified and new inputs for the next simulation are estimated under the assumption, that the system is mostly linear between the seeds.

The algorithm is passed multiple times. One iteration is called an epoch. The initial conditions $x_0$ are kept the same for all simulations in one epoch.

An epoch is composed of four phases. Phase 1 uses classical sampling methods to create the basis for the following adaptive part. Phase 2 expands the convex hull of the seeds in the output space, while phase 3 populates empty regions inside the hull. Phase 4 creates a new set of initial conditions for the next epoch. In the following, each of these phases are detailed further and the settings and termination of the algorithm are discussed.

### 3.1. Phase 1 - Initial Sampling

Phase 1 creates the basis for the adaptive sampling. The input space is a hypercube of dimension $d_i$. Hammersley sequence sampling is used to create samples for $\bar{u}$, which are well distributed in the input domain. Additional samples are set directly on the corners and the center of the faces of this hypercube (see Figure 3a).

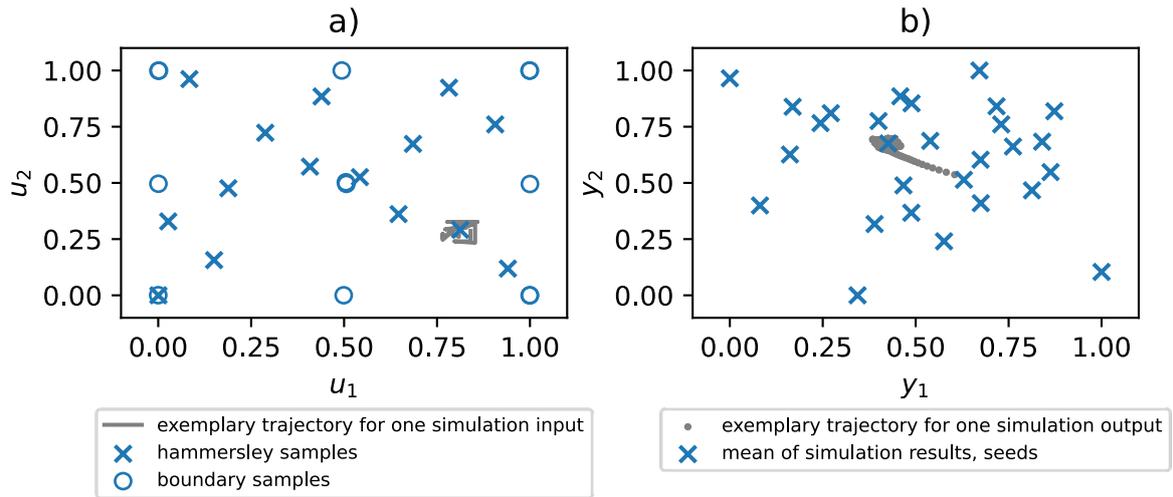

Figure 3. a) Projection of Three-dimensional input space with mean of inputs. b) Seeds of all simulations in output space.

### 3.2. Phase 2 - Expansion

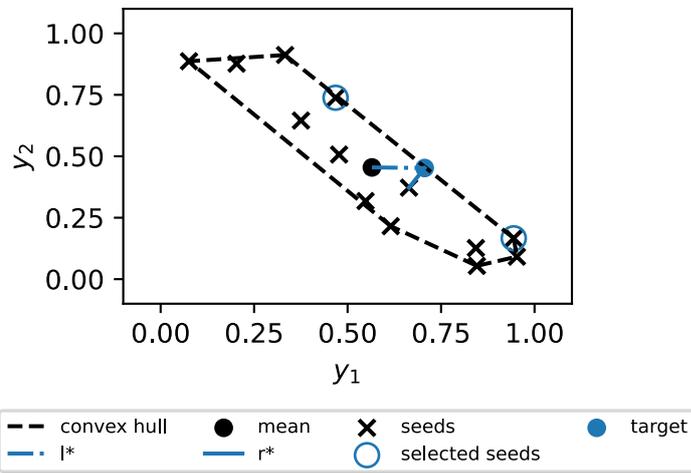

Figure 4. Two-dimensional output space with an exemplary target, the corresponding seeds and necessary values $l^*, r^*$ for calculation of the scoring function.

The goal of phase 2 is to increase the coverage of the output space $Y$, specifically to extend the convex hull of the seeds to cover a larger space. The seed of one simulation is calculated by taking the weighted mean of all simulation results (

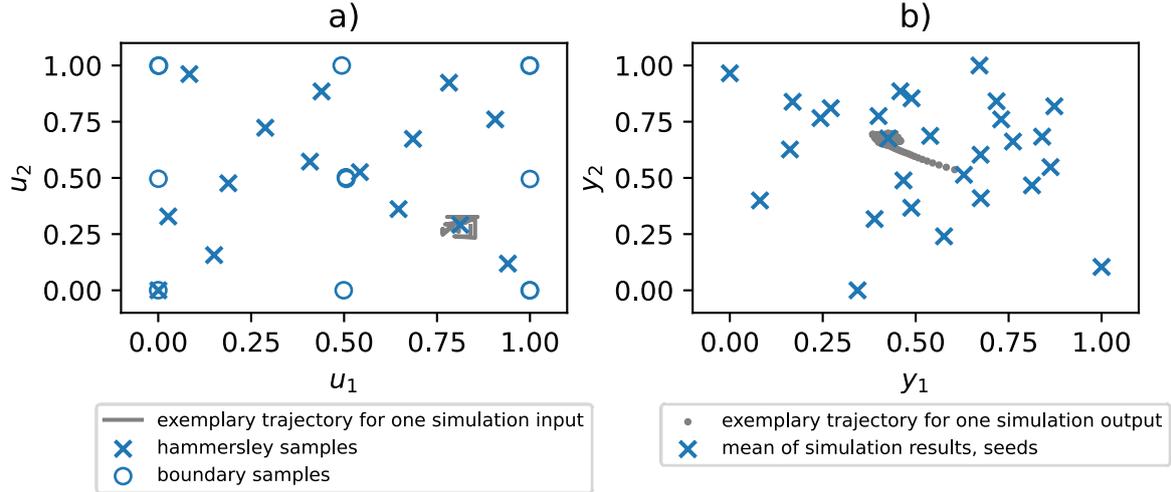

Figure 3b). To achieve this, possible candidates (a new input) and targets (expected value in the output space) are computed. The targets are designed to be close to the current perimeter of the hull and as far away as possible from the seeds. They are scored accordingly. The best candidate gets selected, and the simulation is started.

A candidate consists of input and target and is created by combining exactly two previously run experiments. According to the linearity assumption the input of the candidate is $\bar{u}^* = (\bar{u}_1 + \bar{u}_2)/2$ and the target value in y can be determined as $t^* = (\bar{y}_1 + \bar{y}_2)/2$. All combinatorial possible candidates are calculated and scored.

For scoring the center point of all seeds, M, is computed, and for every target $t^*$ the euclidean distance to M, $l^*$, and to the closest seed, $r^*$, are calculated.

$$l^* = \|t^* - M\|_2 \tag{2}$$

$$r^* = \min\left(\|t^* - \bar{y}_1\|_2, \ldots, \|t^* - \bar{y}_n\|_2\right) \tag{3}$$

One example is shown in Figure 4. All possible targets are then scored: $s^* = f(l^*, r^*)$. $f$ is chosen in such a way, that the score improves for larger $l^*$ and larger $r^*$. To prevent an infinite loop, targets are declared invalid, if they are too close to any previously used target: $\|t^* - t_{used,i}\|_2 < r_{used,i}$.

Phase 2 is repeated until there are no more valid targets, the maximum number of simulations in phase 2 is reached, or a threshold for the scoring function is surpassed. The latter two are hyperparameters for this phase.

### 3.3. Phase 3 – Population

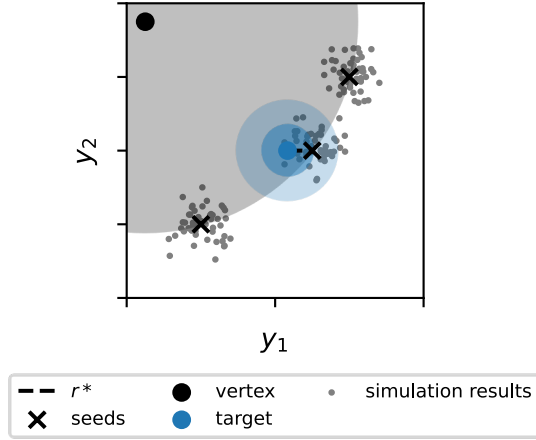

*Figure 5. Phase 3 candidate selection. The vertex defines the selection of seeds, for target calculation. Size and number of simulation results inside the blue n-balls are used in the scoring function.*

The goal of phase 3 is to populate empty regions inside the convex hull of the seeds in the output space. Identifying these empty regions is equivalent to the *largest empty sphere problem*, which is known in computational geometry and can be solved using Voronoi diagrams [5]. A Voronoi-algorithm returns vertices [4], which are the center of spheres defined by the closest seeds and can be used in higher dimensions.

Applying the algorithm on the seeds off the previously run simulations, every vertex defines a set of d+1 experiments. The number of vertices and the computational cost of the algorithm ($\mathcal{O}(n_{simulations\ in\ epoch}^{\lfloor \dim(Y)/2 \rfloor})$) is small in comparison to an exhaustive search ($\mathcal{O}(n_{simulations\ in\ epoch}^{\dim(Y)})$).

For every vertex, a candidate is computed and scored. The criterion is based on the size of spheres surrounding the targets and the number of simulation results - y(t) - inside of them, favoring big spheres with few points inside of them.

Candidates and targets are computed similarly to phase 2, by combining d+1 experiments.

$$\overline{u_i^*} = \frac{1}{d+1} \sum_{j=1}^{d+1} \overline{u_j} \tag{4}$$

$$t_i^* = \frac{1}{d+1} \sum_{j=1}^{d+1} \overline{y_j} \tag{5}$$

A radius is defined as the smallest distance between the target and the defining seeds.

$$r_i^* = \min(\|t_i^* - \overline{y_1^*}\|_2, \dots, \|t_i^* - \overline{y_{d+1}^*}\|_2) \tag{6}$$

The target is scored by the function $s^* = f_\kappa(r^*, n^*)$, wherein $n^*$ describes the number of simulation results inside the d-ball centered at $t^*$. The original outputs $y$ with fixed timesteps are used for counting the simulation results inside a d-ball. $\kappa$ is a hyperparameter, which defines the number of d-balls that are considered (see Figure 5). Especially in higher dimensions the d-ball with radius $1 \cdot r^*$ often is empty, so multiple d-balls with radius $r = 1 \cdot r^*, \dots, \kappa \cdot r^*$ are evaluated. The inner shells have a bigger influence on the scoring function.

The score improves for big radii and small number of results inside the d-balls. To prevent an infinite loop, targets which are close to already used ones, are declared invalid and are not evaluated further.

Phase 3 is repeated until there are no more valid targets left or the maximum number of experiments is reached. During phase 3, as the empty regions are filled, the mean of the computed radii $r^*$ decreases. This serves as an additional termination criterion. The maximum deviation for the mean radius and the number of iterations below that value are hyperparameters as well as the number of evaluated n-balls $\kappa$.

### 3.4. Phase 4 – Restart

If the maximum number of epochs is not reached, a new set of initial conditions for the next epoch is determined with the intention to expand the covered region in output space $Y$. Selecting new initial conditions for a DAE-system is non-trivial. By taking a point from a formerly traversed trajectory it can be guaranteed that the selected point is a valid initialization of the system.

The new initial condition is computed by using all simulations from all epochs. To overcome the issue of the curse of dimensionality, a subset of all state variables must be selected, that is considered further. The center of all results is calculated and the point with the largest distance to the center is selected as new initial condition. A minimum distance to all previously used initial conditions must be maintained. It is proposed to use the average distance between two random points in a hypercube as minimum distance, but it can be chosen freely (24).

The algorithm terminates when there are no more valid initial conditions, or the maximum number of epochs is reached.

### 3.5. Computational complexity

The main influencing factors for each phase are stated below:

The number of simulations in phase 1 depends on the dimensionality of the input space and the chosen number of Hammersley samples.

$$n^{PI} = 2^{d_u} + 2d_u + n_{HSS}^{PI} \tag{7}$$

The number of candidates for each iteration in phase 2 is $\binom{n_{sim,epoch}}{2} = \mathcal{O}(n_{sim,epoch}^2)$, wherein $n_{sim,epoch}$ is the number of simulations in the current epoch, which have to be evaluated.

In phase 3, for eqach iteration the most expensive operation is to calculate and evaluate the matrix of Euclidean distances between the targets and the simulation results. The Voronoi algorithm returns $\mathcal{O}\left(n_{sim,epoch}^{\left\lfloor \frac{\dim(Y)}{2} \right\rfloor}\right)$ vertices and therefore targets. With a fixed timestep and time horizon for all simulations, there are $n_y = n_{sim,epoch} \cdot n_{\Delta t/sim}$ simulation results. So, the matrix is of size $n_{sim,epoch}^{\left\lfloor \frac{\dim(Y)}{2} \right\rfloor} \times n_y$ and must be evaluated for every considered radius for a total $\kappa$ times.

In phase 4 the distance matrix of size $n_{epochs} \times n_{sim,total} \cdot n_{\Delta t/sim}$ must be computed once.

## 4. Case Study

To demonstrate the performance and the applicability for dynamic data-driven modeling, the presented adaptive sampling framework is applied on a dynamic model of a chlor-alkali electrolysis (CAE) and a recurrent neural network is trained and tested based on the generated dynamic data sets.

### 4.1. Problem Description

The chlor-alkali electrolysis produces chlorine, hydrogen and caustic soda for sodium chloride brine using electrical power. A flowsheet of the modeled process is shown in Figure 6a. Here, the CAE cell is represented as a coupled system of two continuously stirred-tank reactors. For a detailed description of the used model, the reader is referred to (25).

The control variables $u$ used for the case study are the current density $j$ applied to the CAE cell, the inlet temperature of the catholyte feed $T_{in}$ and the volume feed flow of the sodium chloride brine $V_{in}$. To manipulate the two latter controls, the two controllers marked in dashed lines in Figure 6a had to be removed from the original model. The lower and upper bounds of $u$ as well as the maximum possible control changes in one time step (amplitude of the FAPRBS) used in the sampling algorithm are listed in Table 1.

*Table 1. Specification of the used FAPRBS control sampling.*

| Control $u$ | Lower bound of $u$ | Upper bound of $u$ | Amplitude of FAPRBS |
|---|---|---|---|
| $j$ in A/m$^2$ | 5000 | 6000 | 200 |
| $T_{in}$ in °C | 59 | 89 | 6 |
| $V_{in}$ in l/s | 0.05 | 0.07 | 0.004 |

The variables that are supposed to be described in the dynamic surrogate model (output space $Y$) are the temperature in the CAE cell $T_{cell}$ and the sodium ion mass fraction in the anolyte $w_{Na^+}$. Both variables are controlled variables of the removed controllers (marked dashed in Figure 6a).

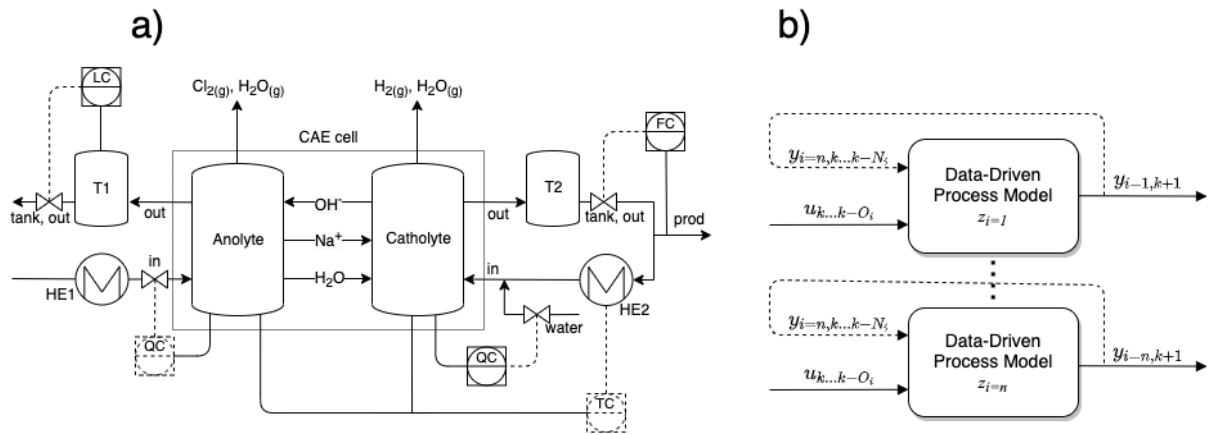

*Figure 6. Model overview: a) Flowchart of chlor-alkali process model, dashed controllers are removed from model and associated manipulated variables are used as input variables in sampling algorithm. b) Structure of used recurrent neural networks. Each output is modeled separately. Parameters N and O are determined in hyperparameter tuning.*

### 4.2. Adaptive Sampling

The presented framework was applied on the CAE system described above. The algorithm finished using 5 epochs (initial conditions) and performed 145, 15 and 121 dynamic simulations in the phases 1, 2 and 3, respectively. Each dynamic simulation used a FAPRBS signal with 30 samples at a frequency of 1000 $s^{-1}$ and 10 samples at a frequency of 2000 $s^{-1}$. The FAPRBS's amplitude specifications are listed in Table 1.

The resulting dynamic samples in the in- and output spaces are shown in Figure 7. Since both output variables used in the algorithm are algebraic variables in the CAE model, the initial results at $t_0$ are distributed over four areas, each corresponding to an initial condition. 97.5% of the computation time was used for the simulations, with the rest spent on the algorithm. Here, calculation and evaluation of the matrix in phase 3 took 81.5% of the computing time, determination of all input signals 14.2%, and the calculation of targets 2.5%. All other subroutines can be neglected with a maximum time usage of less than 0.5% each.

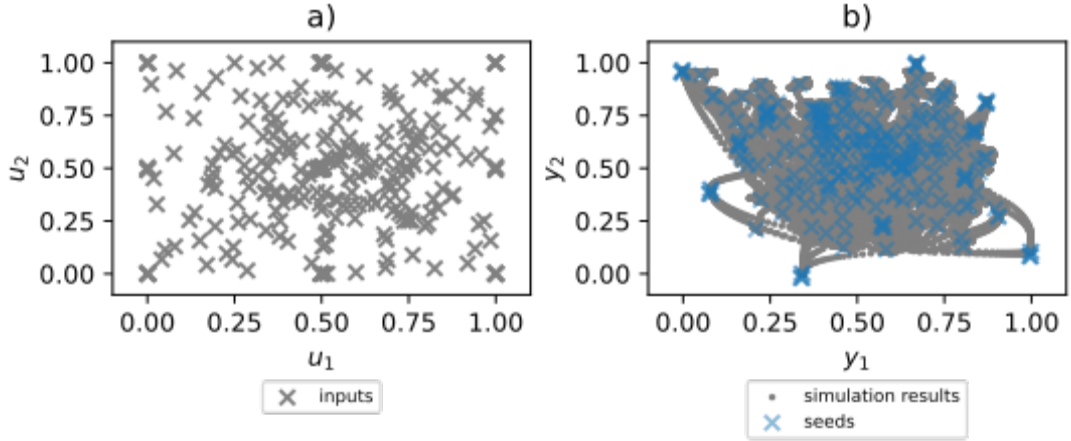

Figure 7. Results of the adaptive sampling algorithm for the CAE model. a) 2-dimensional control space representation for j ($u_1$) and $T_{in}$ ($u_2$). b) seeds and simulation results of the output space.

### 4.3. Dynamic Data-Driven Modeling

To model the dynamic behavior of the predefined output variables $y$, a recurrent neural network was trained for each output separately. The in- and output specifications of the used recurrent neural network are shown in Figure 6b. To predict $y$ at time point $t_{k+1}$ the last $O$ control variable values at the time points $t_{k-O,...,}t_k$ and the last N values of the modeled output variable $y$ at the time points $t_{k-N,...,}t_k$ are fed into the recurrent neural network as input variables.

To find a suitable parameterization of the neural networks a hyperparameter tuning using Bayesian optimization is performed in addition to the standard model training. The varied hyperparameters and the results of the tuning are listed in Table 2.

To test the quality of the resulting models, an additional test set consisting of dynamic data of 5 simulations is used. The testing control variables are again sampled from an FAPRBS using the same specifications as in the adaptive sampling (see Table 1) but with mean control values $\bar{u}$ that were not used in the training data.

Table 2. Parameters and results of hyperparameter tuning using Bayesian optimization.

| Model | Nodes hidden layer | L2 penalty parameter | N | O of j | O of $T_{in}$ | O of $V_{in}$ |
|---|---|---|---|---|---|---|
| $T_{cell}$ | 119 | 0.046 | 15 | 12 | 21 | 8 |
| $w_{Na^+}$ | 242 | 0.149 | 19 | 24 | 7 | 21 |

The standard model training is performed using Adaptive Moment Estimation (Adam) (26). The trained models of the cell temperature and the anolyte composition show a mean squared error regarding the testing data of $4.62 \cdot 10^{-6}$ and $5.36 \cdot 10^{-7}$ (in a normalized output space between 0 and 1), respectively.

Figure 8 shows the testing results of both modeled variables. It can be seen that the dynamic behavior of both variables can be predicted with a high degree of accuracy over a wide value range in the output space. This behavior indicates that the data generated using the presented adaptive sampling algorithm, provides sufficient information over the entire feasible area of the output variables of interest. The comparison with a conventional method for dynamic system identification, which uses an APRBS sampling with an amplitude between the lower and upper bounds of the defined controls (see Table 1 ), could not be carried out, since the simulation did not converge at such large changes.

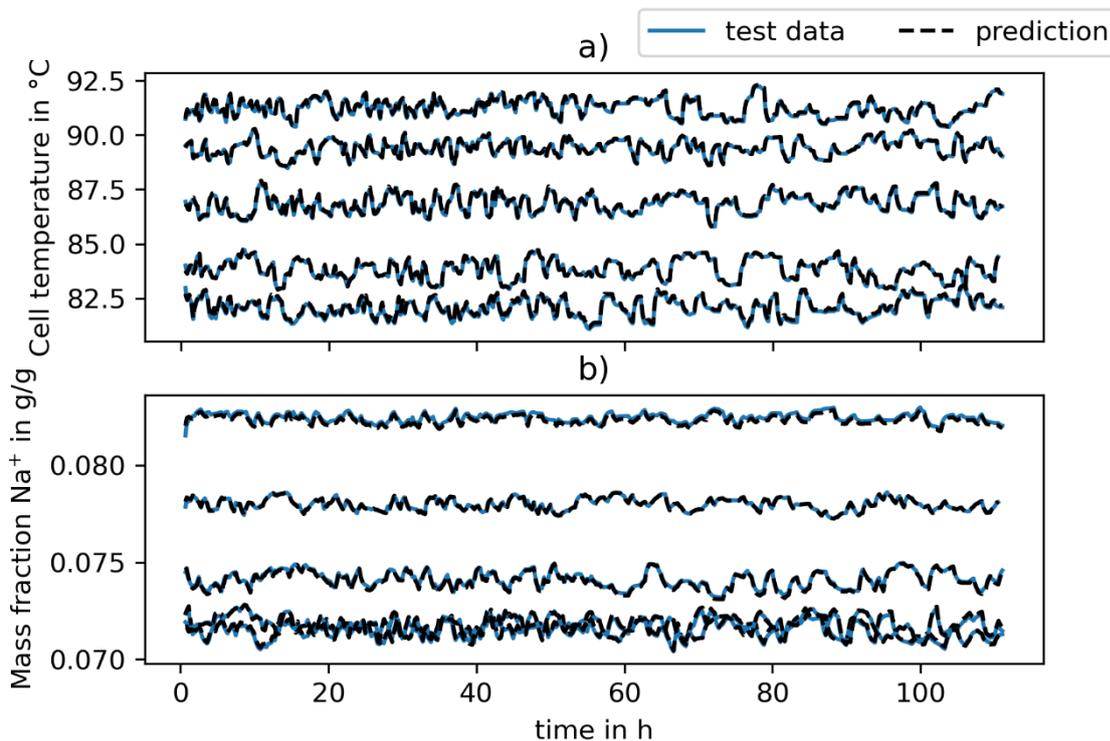

*Figure 8. Results of comparison between test data and model prediction for 5 simulations over 110 hours: a) Cell temperature. b) Mass fraction of sodium ions in anolyte.*

## 5. Conclusion & Outlook

A novel methodology to adaptively sample rigorous dynamic process models to generate a dataset for building a surrogate model is presented. The goal of the developed algorithm is to cover an as large as possible area of the feasible region of the original model. To do so multiple simulations with a short time horizon, a fixed timestep, and different inputs $\bar{u}$ are carried out. In order to maximize the dynamic information of the simulation results the here proposed FAPRBS sampling is used to generate a dynamic trajectory for the different inputs. In the course of the algorithm, empty areas in the output space are identified and the corresponding values in the input space are estimated in order to generate new data in the required area.

To demonstrate the performance and the applicability for dynamic data-driven modeling, the presented framework is applied on a dynamic model of a chlor-alkali electrolysis. It can be shown that the generated data is sufficient for training highly accurate recurrent neural networks for describing the dynamic behavior of the defined output variables over the entire feasible region.

In future work, we will focus on developing techniques to estimate the uncertainty of the trained recurrent neural networks to directly identify areas in the input space where additional data is required.

**Nomenclature**

$d\_u$: number of control variables / dimensionality of input space

$d$: number of selected output variables / dimensionality of output space

$\tau$: time

U input space, control variables

$\bar{u}$: mean of APRBS

X state variables

Y output space

$\bar{y}$: seed

t: target

s: score

r: radius

l: distance

n: number of results in d-ball

$\kappa$: number of considered d-balls